\documentclass[journal]{IEEEtran}

\ifCLASSINFOpdf
\else
\usepackage[dvips]{graphicx}
\fi
\usepackage{url}
\usepackage{balance}
\usepackage{graphicx}
\usepackage{amsfonts}
\usepackage{amsbsy}
\usepackage{amssymb}
\usepackage{times}
\usepackage{graphicx}
\usepackage{enumerate}
\usepackage[usenames]{color}
\usepackage[dvips]{pstcol}
\usepackage{epstopdf}
\usepackage[caption=false,labelfont=rm,textfont=rm]{subfig}
\usepackage{cite}
\usepackage{amssymb}
\usepackage{amsfonts}
\usepackage{graphicx}
\usepackage{epsfig}
\usepackage{psfrag}
\usepackage{xcolor}
\usepackage{amsfonts, bm}
\usepackage{epstopdf}
\usepackage{cite}
\usepackage{color}
\usepackage{xcolor}
\usepackage{verbatim}
\usepackage{multirow}
\usepackage{array}
\usepackage{booktabs}
\usepackage{amsthm}
\usepackage{makecell}
\usepackage{mathrsfs}
\usepackage[linesnumbered, ruled]{algorithm2e}
\usepackage{algpseudocode}
\usepackage{amsmath}

\newtheorem{proposition}{Proposition}

\definecolor{ccr}{RGB}{0,0,255}  
\definecolor{ccb}{RGB}{0,0,255}  
\usepackage{hyperref}
\hypersetup{hypertex=true,
	colorlinks=true,
	linkcolor=ccr,
	anchorcolor=ccr,
	citecolor=ccb}
\IEEEoverridecommandlockouts

\begin{document}

	\title{{Quantize-Sample-and-Verify: LLM Acceleration via Adaptive Edge-Cloud Speculative Decoding}}
	
	\author{{Guangyi Zhang, \textit{Graduate Student Member, IEEE}, Yunlong Cai, \textit{Senior Member, IEEE}, Guanding Yu, \textit{Senior Member, IEEE}, Petar Popovski, \textit{Fellow, IEEE}, Osvaldo Simeone, \textit{Fellow, IEEE}}\vspace{-2.4em}
		\thanks{
			The work of Guangyi Zhang and Yunlong Cai was supported in part by the National Natural Science Foundation of China under Grant 62571477, and in part by Zhejiang Provincial Key Laboratory of Multi-Modal Communication Networks and Intelligent Information Processing, Hangzhou 310027, China. The work of Petar Popovski was supported in part by the Velux Foundation, Denmark, through the Villum Investigator Grant WATER, nr. 37793. The work of Osvaldo Simeone was supported by the European Research Council (ERC) under the European Union’s Horizon Europe Programme (grant agreement No. 101198347), by an Open Fellowship of the EPSRC (EP/W024101/1), and by the EPSRC project (EP/X011852/1). (\emph{Corresponding author: Yunlong Cai})
			
			G. Zhang, Y. Cai, and G. Yu are with the College of Information Science and Electronic Engineering and Zhejiang Provincial Key Laboratory of Multi-Modal Communication Networks and Intelligent Information Processing, Zhejiang University, Hangzhou 310027, China (email: \{zhangguangyi, ylcai, yuguanding\}@zju.edu.cn).
			
			P. Popovski is with the Connectivity Section, Aalborg University, Aalborg 9220, Denmark. (e-mail: petarp@es.aau.dk)
			
			O. Simeone is with the Intelligent Networked Systems Institute (INSI), Northeastern University London, One Portsoken Street, London E1 8PH, United Kingdom (email: o.simeone@northeastern.edu).}}
	\maketitle

	\IEEEpeerreviewmaketitle
	
	\begin{abstract}
		In edge-cloud speculative decoding (SD), edge devices equipped with small language models (SLMs) generate draft tokens that are verified by large language models (LLMs) in the cloud. A key bottleneck in such systems is the limited communication bandwidth between edge and cloud, which necessitates quantization of the information transmitted about generated tokens. In this work, we introduce a novel quantize-sample (Q-S) strategy that provably preserves the output distribution of the cloud-based model, ensuring that the verified tokens match the distribution of those that would have been generated directly by the LLM. We develop a throughput model for edge-cloud SD that explicitly accounts for communication latency. Leveraging this model, we propose an adaptive mechanism that optimizes token throughput by dynamically adjusting the draft length and quantization precision in response to both semantic uncertainty and channel conditions. Simulations demonstrate that the proposed Q-S approach significantly improves decoding efficiency in realistic edge-cloud deployment scenarios.
	\end{abstract}
	

	\section{Introduction}
\begin{figure}[t]
		\begin{centering}
			\includegraphics[width=0.3 \textwidth]{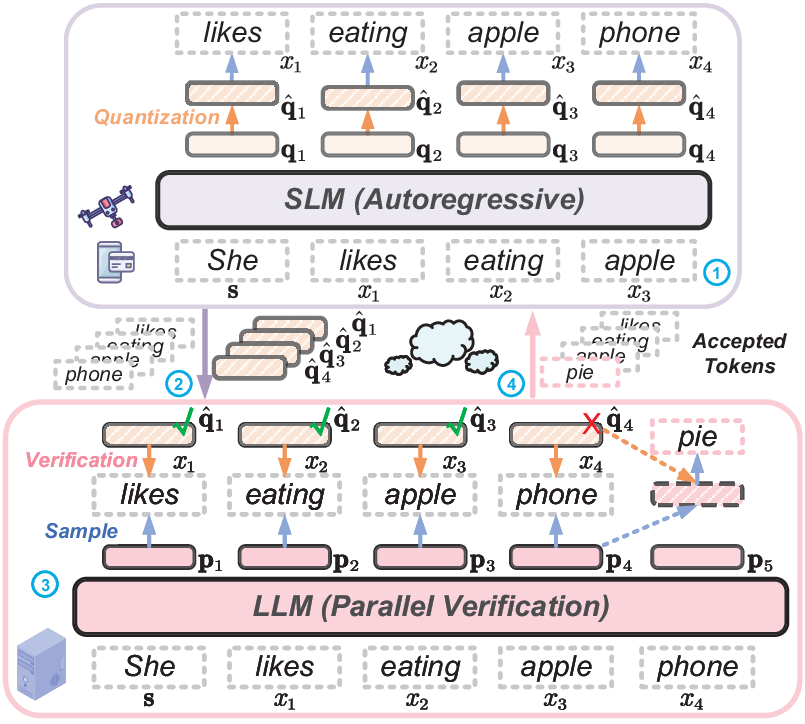}
			\par \end{centering}
		\caption{Schematic illustration of the proposed quantize-sample (Q-S) scheme for a cloud-edge speculative decoding (SD) system operating over a wireless network. The system operates in each iteration $t$ via the following steps: \textcircled{1} token generation, \textcircled{2} uplink transmission, \textcircled{3} token verification, and \textcircled{4} downlink transmission.  ($L^t=4$, with superscript $t$ omitted for clarity.) }
		\vspace{-1.2em}
		\label{Fig1}
	\end{figure}

	As illustrated in Fig.~\ref{Fig1}, edge-cloud \textit{speculative decoding} (SD)  employs a \textit{small language model} (SLM) at the edge to generate multiple draft tokens, which are subsequently verified in parallel by a \textit{large language model} (LLM) in the cloud \cite{Chen2023AcceleratingLL}. This setup exploits the low-latency benefits of parallel verification over traditional autoregressive generation, significantly improving token generation throughput\cite{shaoFlow ,panda_speculativeedgecloud,Zixu_HLM,uncertainty}. 
	Compared with cloud-only SD, edge-cloud SD can provide continuous service to end users even in the presence of cloud outages \cite{s2023self}, and it can significantly reduce operational costs by decreasing redundant API calls \cite{fastinference}.
	
	In edge-cloud deployments\cite{Yicen_ICL}, however, SD necessitates frequent communication between the SLM and the LLM, including the transmission of draft token sequences and their corresponding probability distributions. When operating over constrained or time-varying uplink channels, these communication demands can introduce nontrivial delays that hinder overall decoding performance.

	Edge–cloud collaborative SD was initially explored in \cite{Zixu_HLM}, which proposed to leverage a lightweight on-device SLM to draft tokens and identify uncertain tokens for cloud-based verification by an LLM. To further reduce uplink transmissions to the cloud, the authors of the work \cite{uncertainty} introduced an uncertainty-based skipping strategy, where tokens with high confidence are excluded from verification. In \cite{shaoFlow}, the authors provided a case study about using SD on an image captioning task served at the edge. 
	%A very recent work \cite{Jinwoo_Park} proposed a proactive edge drafting strategy that overlaps edge token generation with server-side verification. 
	In parallel, the early-exit mechanism was integrated into edge-cloud SD systems, enabling the client to preemptively draft subsequent tokens before final verification \cite{panda_speculativeedgecloud}.

	In prior works \cite{shaoFlow,  panda_speculativeedgecloud}, the token distributions were assumed to be transmitted losslessly from edge to cloud. To reduce the communication overhead, the papers \cite{Zixu_HLM} and \cite{uncertainty} proposed to quantize the token distributions \emph{after} the sampling step at the edge. For such \emph{sample-quantize} (S-Q) schemes, the distribution of the tokens generated at the edge deviates from the token distribution used for verification at the cloud. As a result, S-Q SD violates the key property of SD of preserving the cloud LLM token distribution. 
	
	In this context, this work introduces a \emph{quantize-sample} (Q-S) protocol for edge-cloud SD that strictly aligns the output distribution of edge-cloud SD with that of the LLM. Furthermore, rather than assuming negligible communication latency as in prior work, this letter explicitly models both uplink and downlink transmission delays, capturing the effects of dynamic transmission conditions. Based on this analysis, we introduce an adaptive resource allocation mechanism that dynamically tunes draft length and quantization precision based on both semantic uncertainty and current channel conditions.
	
	The rest of this letter is organized as follows. Section II describes the system model, with a focus on the edge-cloud SD procedure. Section III presents the proposed Q-S SD method. In Section IV, we introduce the reinforcement learning-based policy design. Simulation results are provided in Section V, and Section VI concludes this letter.

	\section{System Model}
	As illustrated in Fig.~\ref{Fig1}, we consider an edge computing setup comprising an edge device and a cloud equipped with a high-performance computational server. The prompts are sent to the edge device, where an SLM is deployed, while an LLM runs on the server. In a conventional mobile edge computing protocol, all prompt-related packets are uploaded to the cloud for processing by the LLM using the uplink, and the generated tokens are then fed back to the edge device on the downlink. This scheme typically experiences a low end-to-end token throughput due to the computing latency associated with autoregressive generation by the LLM. To increase the end-to-end throughput, edge-cloud SD operates iteratively, with each iteration $t$ consisting of four phases: \textcircled{1} token generation on the device; \textcircled{2} uplink transmission from the device to the server; \textcircled{3} token verification on the server; and \textcircled{4} downlink transmission from the server to the device.
	
	\subsubsection{{Token Generation}}
	At each iteration $t$, the device uses the local SLM to sample $L^t$ draft tokens in an autoregressive manner. To this end, the SLM takes as input the current \emph{prefix} $\mathbf{s}^t=\left[s_1,s_2,...,s_{M^t}\right]^T$, whose $M^t$ tokens encompass the user prompt and all tokens generated so far.  Denote as $x_l^t \in \{1,2,...,V\}$ the $l$-th token produced at iteration $t$, for all $l=1,2,...,L^t$, where $V$ denotes the vocabulary size. Each token $x_l^t$ is generated by using the $V\times 1$ probability distribution, $\mathbf{q}_l^t =[{q}_{l,1}^t,{q}_{l,2}^t,..., {q}_{l,V}^t]^T$, which is evaluated by the SLM by processing autoregressively all previously generated tokens. 
	
	Accordingly, at each iteration $t$, the SLM generates the block of $L^t$ tokens
	$\mathbf{x}^t = [x_1^t, x_2^t, ..., x_{L^t}^t]^T$,
	and we write
	$
	\mathbf{Q}^t=\left[\mathbf{q}_1^t,\mathbf{q}_2^t,...,\mathbf{q}_{L^t}^t  \right]
	$
	for the collection of all probabilities produced by the SLM at iteration $t$.
	
	\subsubsection{{Uplink Transmission}}
	Using uplink transmissions, at each iteration $t$, the device transmits information to the server about the tokens $\mathbf{x}^t$ and the probability vectors $\mathbf{Q}^t$. Communications from device to the server during iteration $t$ is constrained by an uplink rate $C^t_{\mathrm{u}}$ (bits/s).  
	
	Each token requires $\lceil{\log_2 (V)}\rceil$ bits to be represented losslessly, while the number of bits used to quantize each probability vector $q^t_l$ for $l=1,2,...,L^t$ is denoted as $b^t$. We denote the quantized probability vectors as
	$
	\mathbf{\hat{Q}}^t=[\mathbf{\hat{q}}_1^t,\mathbf{\hat{q}}_2^t,...,\mathbf{\hat{q}}_{L^t}^t  ]
	$, where the $l$-th probability vector is expressed as $\mathbf{\hat{q}}_l^t=[{\hat{q}}_{l,1}^t,{\hat{q}}_{l,2}^t,..., {\hat{q}}_{l,V}^t]^T$.
	
	\subsubsection{{Token Verification}}
	After receiving the token sequence $\mathbf{x}^t$ and probabilities $\mathbf{\hat{Q}}^t$, the server employs an LLM for token verification. Note that, by design, the server is aware of all the previously generated tokens, and thus it has access to the prefix $\mathbf{s}^t$. Using the prefix and the received token sequence $\mathbf{x}^t$ as inputs, the server can produce the $L^t+1$ probability vectors \( \mathbf{p}_1^t, \mathbf{p}_2^t, \dots, \mathbf{p}_{L^t}^t, \mathbf{p}_{L^t+1}^t \) in parallel, where each probability vector $\mathbf{p}_l^t =[{p}_{l,1}^t,{p}_{l,2}^t,..., {p}_{l,V}^t]^T$ depends on the prefix $\mathbf{s}^t$ and on the tokens $[x_1^t,x_2^t,...,x_{l-1}^t]^T$. The server then uses the probability vector $\mathbf{p}_{l}^t$ to decide whether to accept or reject the $l$-th token $x_l^t$.
	
	In particular, starting with $l=1$ and proceeding sequentially,  the server accepts token $ x_l^t$ with probability \cite{fastinference}
	\begin{equation}\label{Pro2}
		\alpha_l^t=	\min \left( 1, \frac{{p}_{l,x_l^t}^t}{{\hat{q}}_{l,x_l^t}^t} \right)
	\end{equation}
	for \( l = 1, 2, \dots, L^t \). If token \( x_l^t \) is the first to be rejected, a new token is sampled as
	\begin{equation}\label{Pro3}
		\tilde{x}_l^t \sim  \mathbf{\tilde{p}}_l^t = \frac{\max \left( 0, \mathbf{p}_l^t - \mathbf{\hat{q}}_l^t \right)}{\sum_{j=1}^{V} \max \left( 0, {p}_{l,x_j^t}^t - {\hat{q}}_{l,x_j^t}^t \right)},
	\end{equation}
	where the $\max(\cdot,\cdot)$ operation is applied element-wise.
	Alternatively, if all $L^t$ tokens are accepted, a new token $\tilde{x}^t_{L^t+1}$ is sampled from the distribution. We denote as $N^t \leq L^t$ the number of tokens accepted by the server. 
	
	\subsubsection{{Downlink Transmission}}
	After the verification phase, the device needs to be notified about $N^t$, the number of tokens accepted by the server, as well as about the newly generated token at the server, namely $\tilde{x}^t_{l}$ or $\tilde{x}^t_{L^t+1}$. This requires the transmission of $\lceil\log_2 (L^t)\rceil+\lceil{\log_2 (V)}\rceil$ bits, where $\lceil\log_2 (L^t)\rceil$ is the number of bits used to describe the integer $N_t$, and $\lceil{\log_2 (V)}\rceil$ is the number of bits required to represent the new token produced by the cloud.
	
	Upon receiving this information, the accepted tokens and the resampled token are appended to the current prefix, i.e.,
	\begin{equation}\label{eq:conc}
		\mathbf{s}^{t+1}= \operatorname{concatenate}(\mathbf{s}^t,x^t_1,x^t_2,...,x^t_{N^t},\tilde{x}^t_{N^t+1}).
	\end{equation}
	This forms the initial prefix for the next iteration $t+1$. The process is stopped when either an end-of-sentence (EoS) token is included in the prefix or a maximum number of tokens is produced \cite{fastinference}.
	
	\section{Quantize-Sample Speculative Decoding}
	
	This section introduces the Q-S protocol, and formalizes its validity property.
	\subsection{Quantize-Sample-and-Verify}
	As discussed in the previous section, in edge-cloud SD, it is necessary to quantize the SLM's output probability vectors $\mathbf{Q}^t$ before transmission to the cloud. Prior art \cite{Zixu_HLM,uncertainty} follows an S-Q approach, whereby draft tokens $\mathbf{x}^t$ are first generated from distributions $\mathbf{Q}^t$ and then the quantized distributions $\hat{\mathbf{Q}}^t$ are obtained. In contrast,  this work proposes a Q-S approach, in which draft tokens $\mathbf{x}^t$ are drawn from the quantized distributions $\hat{\mathbf{Q}}^t$. As we demonstrate next, this design ensures that the generated tokens are statistically equivalent to those that would be generated directly by the LLM.
	
	\begin{figure}[t]
		\begin{centering}
			\includegraphics[width=0.5 \textwidth]{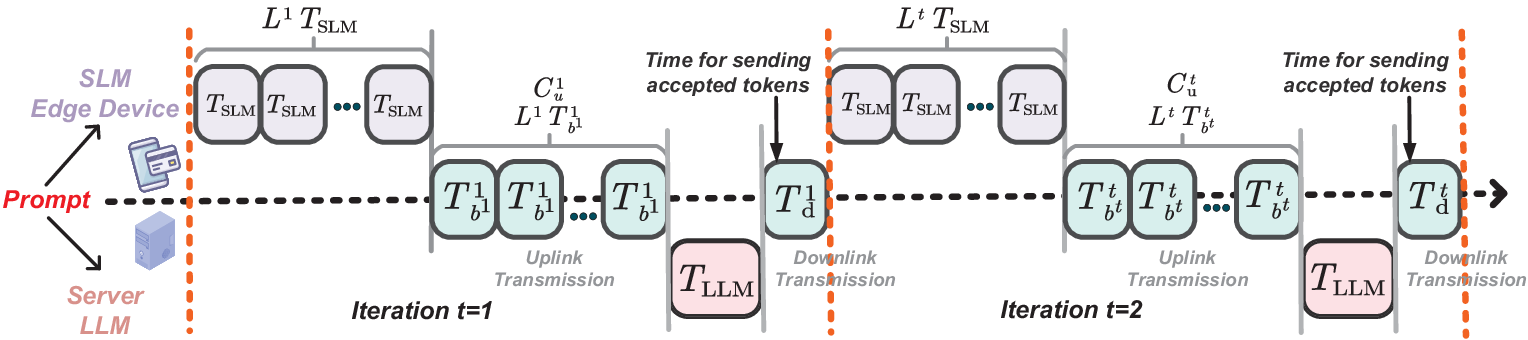}
			\par \end{centering}
		
		\caption{Timeline of the edge–cloud SD operation under the Q-S strategy.}
		\label{Fig2}
		\vspace{-1em}
	\end{figure}
	
	\begin{proposition}\label{proposition1}
		Edge-cloud SD via Q-S guarantees that the probability of generating token $x_l^t$ at iteration $t$, $\mathbb{P}(X = x_l^t)$, coincides with the corresponding probability $p^t_{l,x_l^t}$ for the LLM, i.e.,
		\begin{equation}
			\mathbb{P}(X = x_l^t) = p^t_{l,x_l^t}.
		\end{equation}
	\end{proposition}
	
	This result follows directly from the known properties of SD \cite{fastinference} by viewing the quantized distribution $\hat{\mathbf{Q}}^t$ as the generating probability distribution for the SLM. In fact, the SD protocol accepts and samples tokens, via (\ref{Pro2}) and (\ref{Pro3}), using the quantized distributions. This proposition establishes the equivalence between the outputs of the cloud LLM and of Q-S edge-cloud SD protocol. This result thus demonstrates that employing Q-S SD is, in terms of token quality, equivalent to using the cloud LLM. Consequently, unlike S-Q methods \cite{Zixu_HLM, uncertainty}, the proposed scheme does not introduce any quality degradation. The key reason for this critical difference between S-Q and Q-S is that the former uses two distinct distributions for sampling and verification, while the latter uses the quantized edge model's distribution for both steps.

	\subsection{Lattice-Based Quantization}
	In general, the validity property stated in Proposition \ref{proposition1} holds for any quantization scheme at any resolution. In this paper, we adopt a state-of-the-art quantization algorithm for probability vectors, namely lattice-based quantization \cite{lattice_Ren}. Note that the literature also reports quantization schemes that operate at the level of prompts \cite{jiao2024vector,hao2025quantized} or of the key-value cache \cite{he2024zipcache}. However, these methods are not applicable to the problem of quantizing probability distributions.
	
	Lattice-based quantization restricts probability vectors to a lattice $Q_{\ell}$ that forms a discrete subset of the $V$-dimensional probability simplex. Specifically, we choose a positive integer $\ell$ such that each probability $q_i$ in the vector can be expressed as a rational number with denominator $\ell$, i.e.,  $q_i = o_i/\ell$, where $o_i$ are non-negative integers summing to $\ell$. The set of all such quantized probability vectors of dimension $V$ is defined as\begin{equation}\label{eq:lattice}
		Q_{\ell}=\left\{[q_1,q_2, \ldots, q_V] \in \mathbb{Q}^V \left\lvert\, q_1=\frac{o_1,}{\ell}\right., \sum_{i=1}^{V} o_i=\ell\right\},
	\end{equation}which represents the allowable probabilities after quantization.
	
	To quantize an arbitrary probability distribution, the algorithm finds the nearest lattice point in $Q_{\ell}$ under a given distance metric. As discussed in \cite{Teku_perb}, this operation entails a complexity order of $\mathcal{O}(V\log(V))$, which is negligible compared to the generation cost of SLM. The index of this lattice point can then be encoded using\begin{equation}\label{eq:ell}
		b=\left\lceil\log _2\binom{\ell+V-1}{V-1}\right\rceil
	\end{equation} bits, which is the number of bits needed to identify one point in the probability set $Q_{\ell}$\cite{lattice_Ren}. Accordingly, by fixing the integer $\ell^t$ at iteration $t$  one obtains the corresponding number of bits $b^t$ in (\ref{eq:ell}) with $b^t$ in lieu of $b$.
	
	Lattice-based quantization algorithm has a complexity of $\mathcal{O}(V \log (V))$. In practice, this adds only a delay of a few microseconds per batch on modern GPUs.

	\section{Dynamic Throughput Optimization}
	The objective of SD is to improve the token throughput by reducing inference latency. Accounting for the combination of both, this section first provides a detailed analysis of the latency involved in the edge-cloud SD systems discussed above. Then, it introduces a dynamic policy for throughput optimization.
	
	\subsection{Latency Analysis}
	We denote by $T_{\text{SLM}}$ and $T_{\text{LLM}}$ the time required for the SLM and LLM, respectively, to generate a single token. Typically, one has the strict inequality $T_{\text{SLM}} \ll T_{\text{LLM}}$ since the LLM has far more parameters, leading to higher memory access and computation latency. For example, on an NVIDIA A100 GPU, a large $13$B-parameter model might take around $32$ ms per token, whereas a small $125$M-parameter model needs only about $5$ ms. 
	
	In addition to processing delays, we must account for communication delays between edge and cloud at each iteration $t$. Let $L^t$ be the number of draft tokens generated by the SLM and $b^t$ be the number of quantization bits used for each token’s distribution at iteration $t$. We consider standard packet-level transmission, which reduces protocol overhead as compared to per-token transmission. The uplink transmission time for sending these $L^t$ token probabilities and associated token identities can be modeled as\begin{equation}\label{eq7}
		T_{\mathrm{u}}^t(L^t,b^t;C^t_{\mathrm{u}}) =L^t \frac{ \lceil{\log_2 (V)}\rceil +b^t }{C_{\mathrm{u}}^t},
	\end{equation}where $V$ is the vocabulary size; $\lceil \log_2 (V)\rceil$ accounts for the bits needed to identify each token; $b^t$ is the number of  bits for the quantized probability; and $C^t_{\mathrm{u}}$ is the uplink data rate (in bits/s) at iteration $t$. Similarly, the downlink time to return the verification results for $L^t$ tokens can be expressed as\begin{equation}\label{eq8}
		T_{\mathrm{d}}^t(L^t;C^t_{\mathrm{d}}) = \frac{ \lceil\log_2 (L^t)\rceil+\lceil{\log_2 (V)}\rceil }{C_{\mathrm{d}}^t}, 
	\end{equation}where $C^t_d$ is the downlink rate and $\lceil \log_2 (L^t\rceil)$ accounts for transmitting the number of accepted tokens. In practice, the uplink latency (\ref{eq7}) and downlink latency (\ref{eq8}) also incorporate a fixed overhead due to the transmission of preambles. In the following, we assume these additional terms to be negligible, although the analysis can be readily extended to accommodate them.

	As illustrated in Fig.~\ref{Fig2}, combining the computation and communication components, the total latency for one iteration $t$ is\begin{equation}\label{latency}
		T^t(L^t, b^t;C^t_{\mathrm{u}},C^t_{\mathrm{d}}) = L^t T_{\mathrm{SLM}} + T_{\mathrm{u}}^t+ T_\mathrm{LLM} + T_{\mathrm{d}}^t,
	\end{equation} where $L^t T_{\text{SLM}}$ is the time for the SLM to generate $L^t$ draft tokens (since it generates tokens one by one); $T^t_{\mathrm{u}}$ is the uplink delay to send those drafts; $T_{\text{LLM}}$ is the cloud verification time (since the LLM verifies $L^t$ tokens in parallel); and $T^t_{\mathrm{d}}$ is the downlink time to receive the results.  If the SD process runs for a total of $I$ iterations, the overall decoding latency is the sum over all iterations\begin{equation}
		T(L, b) = \sum_{t=1}^{I}T^t(L^t, b^t;C^t_{\mathrm{u}},C^t_{\mathrm{d}}),
	\end{equation}where we denote $L = [L_1, L_2, \ldots, L_I]$ and $b = [b_1, b_2, \ldots, b_I]$ as the sequences of chosen draft lengths and quantization bits across iterations, respectively.
	
	\subsection{Problem Definition}
	To formally capture the decoding process, let $\mathbf{s}^t$ denote the prefix  after the $t$-th iteration of SD (see (\ref{eq:conc})). We define the stopping set $\mathcal{S}^*$ as containing all completed sequences, i.e., sequences that, when used as a prefix, do not require the generation of any additional token. In particular, we include in the set $\mathcal{S}^*$ any sequence that has reached a predefined maximum length or has produced the EoS token: $\mathcal{S}^*=\{\mathbf{s}^t:|\mathbf{s}^t|\geq N_{\text{max}}+M^1 \; \text{or} \;\text{[EoS]} \in \mathbf{s}^t\}$, where $N_{\max}$ is the maximum allowed sequence length (including the initial prompt). The stopping iteration $\tau^*$ is the first iteration at which the prefix enters this set, i.e., \begin{equation}\label{eq:term}
		\tau^* = \min \{t \geq 1 | \mathbf{s}^t\in \mathcal{S}^*\}.\end{equation}
	
	Let $\mathbf{s}^1$ be the initial prompt (prefix) given to the system. We denote by $\Delta{t} = |\mathbf{s}^{t}|-|\mathbf{s}^{1}|$ the total number of generated tokens at iteration $t$, and thus $\Delta \tau^* = |\mathbf{s}^{\tau^*}| - |\mathbf{s}^1|$ denotes the total number of new tokens generated by the time decoding stops. Our objective is to design a control \emph{policy} $\bm{\pi}$ that maximizes the token throughput, defined as the ratio of new tokens to the total time taken. At the beginning of each iteration $t$, the policy $\bm{\pi}$ maps the current information available at the edge device to the \emph{action} $a^t = (L^t, b^t)$ encompassing  draft length $L^t$  and quantization bits  $b^t$  for the next iteration. 
	In practice, both $L^t$ and $b^t$ are selected from discrete sets $\mathcal{L}$ and $\mathcal{B}$ of allowed values, i.e.,  $L^t \in \mathcal{L}$ and $b^t \in \mathcal{B}$.

	The input \textit{state} $s^t$ to the policy $\bm{\pi}$ at the beginning of iteration $t$ includes: (\emph{i}) \emph{semantic information} in the form of the prefix confidence vector $\mathbf{f}^t = [f_1, f_2, \ldots, f_{\Delta{t}}]$, where each $f_i$ is the probability assigned by the SLM to token $i$ in the current prefix $\mathbf{s}^t$, and of the mean confidence feature
	$\bar{f}^t = \frac{1}{\Delta{t}} \sum_{i=1}^{\Delta{t}} f_i$; and (\emph{ii}) \emph{connectivity information} in the form of the current uplink channel rate $C^t_{\mathrm{u}}$. 
	By incorporating the uplink rate $C_{\mathrm{u}}^t$ into the decision-making process, the agent can adaptively adjust its action policy based on connectivity information. 
	
	Overall, the optimization problem can be written as\begin{equation}\label{p0}
		\begin{aligned}
			\max_{\bm{\pi}} \; \mathbb{E}\left[ \frac{\Delta{\tau_*}}{\sum_{t=1}^{\tau^*} T^t(L^t, b^t;C^t_{\mathrm{u}},C^t_{\mathrm{d}})}\right], \\  
		\end{aligned}
	\end{equation}  subject to the termination condition $\tau^*$ from (\ref{eq:term}). 
	This problem is non-trivial since smaller values of $L^t$ and $b^t$ reduce the per-iteration latencies $T^t$ while also resulting in an increased termination steps $\tau^*$. In fact, fewer draft tokens $L^t$ and a lower resolution $b^t$ imply that more iterations $\tau^*$ are needed to reach the stopping set $\mathcal{S}^*$.

	\subsection{Reinforcement Learning-Based Throughput Optimization}
	To address the problem (\ref{p0}), we adopt a reinforcement learning approach. In particular, we design a reward signal to provide the learning agent with immediate feedback on the quality of its action $a^t$. Ideally, we want the reward to reflect the token throughput achieved. A natural definition of the reward at the terminal iteration $\tau^*$ would be the overall throughput. However, assigning a reward only at the end of the sequence would make learning difficult due to sparse feedback. Instead, we give a non-zero reward at each intermediate iteration that quantifies the contribution of that iteration to the final throughput. 
	Specifically, we define the immediate reward at iteration $t$ as the throughput gained in that iteration, i.e., \begin{equation}
		R(s^t, a^t) = \frac{N_{\text{avg}}^t(L^t, b^t)}{T^t(L^t, b^t; C_{\mathrm{u}}^t, C_{\mathrm{d}}^t)},
	\end{equation}  where $N_{\text{avg}}^t(L^t, b^t)$ is the expected number of newly accepted tokens at iteration $t$, given the current prefix and the action $(L^t, b^t)$. This expectation can be expanded as\begin{equation}
		N_{\text{avg}}^t(L^t, b^t) = \mathbb{E}_{\mathbf{x}^t \sim \hat{q}(\mathbf{x}^t \mid \mathbf{s}^t)} \left[ g(\mathbf{x}^t) \right],
	\end{equation}  where $\mathbf{x}^t = [x_1^t, x_2^t, ..., x_{L^t}^t]^T$ denotes the set of $L^t$ draft tokens sampled by the SLM at iteration $t$, $\hat{q}(\mathbf{x}^t | \mathbf{s}^t)$ is the joint distribution of those tokens under the SLM's (quantized) output, and $g(\mathbf{x}^t)$ is the expected number of tokens in $\mathbf{x}^t$ that are accepted by the LLM. Given that the LLM’s verification process can be modeled as a sequence of Bernoulli trials, the function $g(\mathbf{x}^t)$ can be expressed as $
	g\left(\mathbf{x}^t\right)=\sum_{l=1}^{L^t} l\left(\prod_{j=1}^{l-1} \alpha_j^t\right)\left(1-\alpha_l^t\right)+\left(L^t+1\right) \prod_{j=1}^{L^t} \alpha_j^t.$
	
	We adopt a double deep Q-Network (DDQN) model to implement the policy $\bm{\pi}$ \cite{Guez_Silver_2016}. The Q-network consists of several lightweight linear layers, and is trained offline following the standard procedure in \cite{Guez_Silver_2016}.

	\section{Simulation Setup and Evaluation} 
	\subsection{Setting}
	We adopt a simulation environment modeling a time-varying uplink channel through a discrete-time two-state Markov model. In this model, the uplink alternates between a low-rate state and a high-rate state. The state transitions are governed by transition probabilities between low- and high-capacity states, and vice versa, given by $p_{\textrm{low} \rightarrow \textrm{high}}$ and $p_{\textrm{high} \rightarrow \textrm{low}}$, respectively. We assume delay-free downlink transmission.
	We indirectly control the number of quantization bits per token, $b^t$ by adjusting the integer $\ell^t$ in (\ref{eq:lattice}). A larger integer $\ell^t$ yields a finer probability resolution, and correspondingly a larger number of bits $b^t$. For both LLM and SLM, we can control the \emph{sampling temperature} used for token generation, with a larger temperature providing a larger-entropy output \cite{fastinference}. We evaluate the throughput obtained after the optimization phase.

	We use the OPT-13B language model as the cloud LLM and the smaller OPT-125M model as the edge SLM \cite{zhang2022opt}, targeting the task of abstractive text summarization on the CNN/DailyMail dataset \cite{nallapati}. All experiments are conducted on an NVIDIA A100 40GB GPU for consistency. Inference is performed with a batch size of $1$, which aligns with standard practice in existing literature\cite{Zixu_HLM,uncertainty,shaoFlow,panda_speculativeedgecloud}.
	Rather than directly deploying OPT-125M on an edge device, both models are executed on the A100 GPU to ensure controlled and reproducible measurements. Specifically, we introduce a scaling factor to approximate the runtime difference between the A100 and the representative edge GPUs. The factor is derived from their token generation speed ratio of the given hardware platforms under identical prompts, and the measured latency $T_{\mathrm{SLM}}$ is then divided by this factor to simulate realistic edge-side inference latency. Code is available at this \href{https://github.com/zhang-guangyi/QS-Speculative-Decoding}{link}.
	
	As benchmarks, we consider cloud-only SD, which has zero edge-cloud communication delays, as well as a heuristic dynamic policy following a simple adaptive rule: Starting with a fixed initial draft length $L^1$, the algorithm increases the draft length by $1$ if all generated tokens are accepted, while it sets the next draft length to the number of accepted tokens $N^t$ if only a subset of tokens is accepted. 
	
	\begin{figure}[t]
		\begin{centering}
			\subfloat[]{\label{f1}\includegraphics[width=4.1cm]{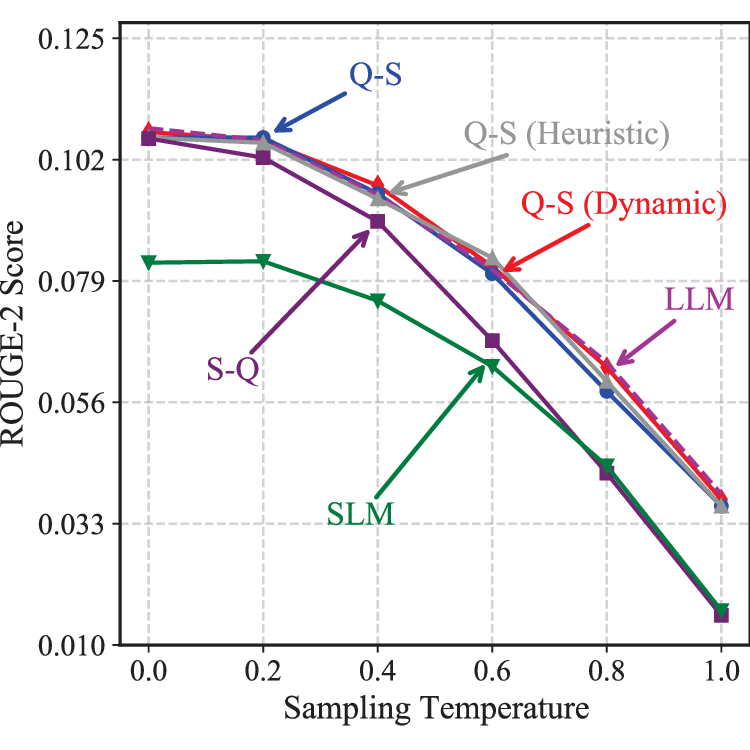}}
			\subfloat[]{\label{f2}\includegraphics[width=4.1cm]{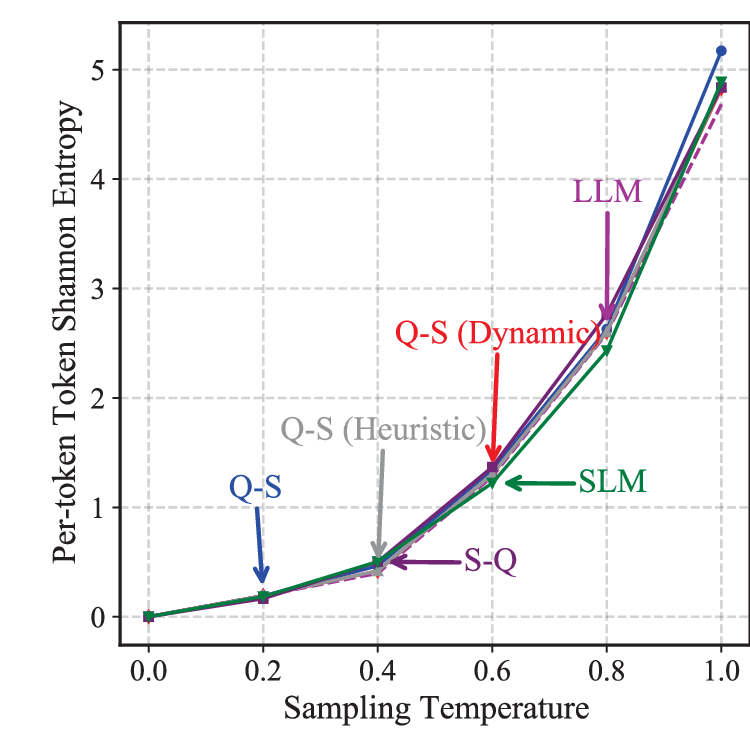}}
			
			\caption{Performance versus sampling temperature for both SLM (OPT-125M) and LLM (OPT-13B): (a) ROUGE-2 score; (b) Per-token Shannon entropy.}
			\label{Fig3}
			\vspace{-1em}
		\end{centering}
	\end{figure}
	
	\begin{figure}[t]
		\begin{centering}
			\subfloat[]{\includegraphics[width=4.2cm]{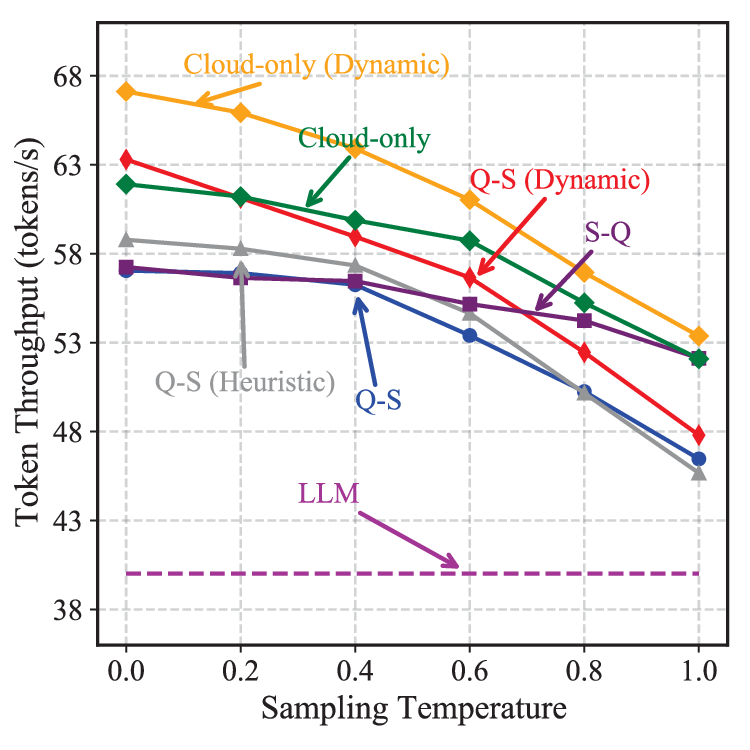}}
			\subfloat[]{\includegraphics[width=4.2cm]{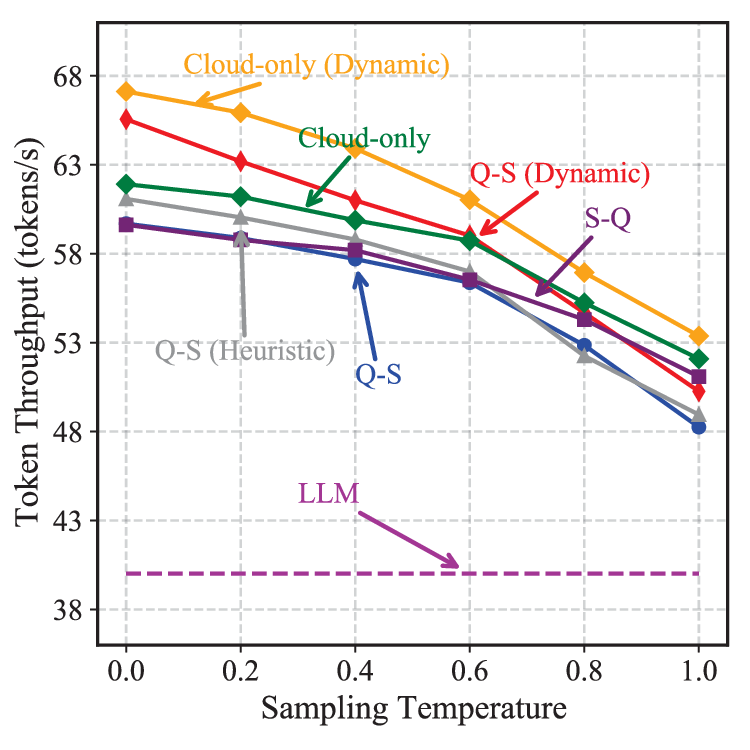}}
			
			\caption{Token throughput achieved by different strategies as a function of sampling temperature. (a) Low-rate regime; (b) High-rate regime.}
			\label{Fig4}
			\vspace{-1em}
		\end{centering}
	\end{figure}
	
	\subsection{Results}
	The left panel of Fig. \ref{Fig3} shows the ROUGE-2 score,  which measures the overlap of bigrams (two-word sequences) between  generated text and reference text from the data set, as a function of the sampling temperatures of both SLM and LLM. However, the right panel plots the token entropy, confirming that higher temperatures introduce greater randomness and output variability \cite{nallapati}.
	
	Among edge-cloud schemes, existing S-Q strategies are seen to fail to match the accuracy of the LLM, particularly at high temperatures. 
	In contrast, the proposed Q-S approach  can maintain the same token quality as the LLM across all sampling temperatures, with or without dynamic optimization. 
	
	Fig. \ref{Fig4} plots the achieved throughput (tokens per second) versus the sampling temperature. Fig. \ref{Fig4}(a) shows the results in a low-rate regime ($350$ kbps on average), which is typical in bandwidth-constrained low-rate systems such as NB-IoT\cite{NB-IOT} and satellite-oriented communications\cite{Wang_Sta}, while Fig. \ref{Fig4}(b) shows the results in a higher rate regime ($4$ Mbps on average). 
	Cloud-only SD clearly outperforms edge-cloud SD, because it incurs no additional communication overhead. Furthermore, as the transmission rate increases, the performance of edge-cloud SD approaches that of cloud-only SD.
	The figure also demonstrates the advantages of edge-cloud SD over LLM in terms of token throughput. As shown in Fig. \ref{Fig4}(a), this benefit comes without loss of accuracy when deploying Q-S, while the use of SD via S-Q entails a reduction in accuracy. 
	
	When using Q-S, Fig. \ref{Fig4} confirms that the proposed dynamic optimization can significantly enhance the throughput across all values of the temperature.
	Specifically, we observe that the dynamic strategy introduced in this work consistently outperforms all static baselines, as well as the heuristic adaptive strategies described earlier across the entire temperature range. 
	
	\section{Conclusion}
	In this letter, we proposed a novel edge-cloud SD strategy that maintains strict consistency with LLM output distributions. To optimize token throughput, we developed a dynamic control mechanism that adapts draft length and quantization precision to both semantic uncertainty and channel conditions based on reinforcement learning. Simulation results validate the efficacy of our approach, showing significant gains in token throughput compared to static methods.

	\bibliographystyle{IEEEtran}
	\bibliography{IEEEabrv,Reference}
\end{document}